# Epitaxial Sc$_x$Al$_{1-x}$N on GaN is a High K Dielectric

Joseph Casamento[1, a)], Hyunjea Lee[2], Takuya Maeda[3], Ved Gund[2],

Kazuki Nomoto[2], Len van Deurzen[4], Amit Lal[2], Huili (Grace) Xing[1,2,3], and Debdeep Jena[1,2,3]

[1]Department of Materials Science and Engineering, Cornell University, Ithaca, NY 14853, USA

[2]School of Electrical & Computer Engineering, Cornell University, Ithaca, NY 14853, USA

[3]Kavli Institute at Cornell for Nanoscale Science, Cornell University, Ithaca NY 14853, USA

[4] School of Applied and Engineering Physics, Cornell University, Ithaca, NY 14853, USA

a) Corresponding author: jac694@cornell.edu

**Abstract:** Epitaxial Sc$_x$Al$_{1-x}$N thin films of ~100 nm thickness grown on metal polar GaN exhibit significantly enhanced relative dielectric permittivity ($\varepsilon_r$) values relative to AlN. $\varepsilon_r$ values of ~17 to 21 for Sc contents of 17 to 25% (x=0.17 to 0.25) measured electrically by capacitance-voltage (CV) measurements at 500 kHz frequency indicate Sc$_x$Al$_{1-x}$N has the largest relative dielectric permittivity of any existing nitride material. This points toward the usage of Sc$_x$Al$_{1-x}$N as potential epitaxial, single-crystalline dielectric material that can be deposited *in situ* on GaN and AlN electronic and photonic devices for enhanced performance.

In condensed matter, other than metals, dielectrics are among the broadest class of materials. With the magnitude of the real part of the relative dielectric permittivity usually exceeding a value of 1, condensed mater materials can store more charge in a finite volume than air and gases can. Dielectrics are utilized in a variety of applications, including usage as passivation layers for optoelectronic devices and as insulating layers between the gate metal and electron and/or hole channels of transistors. A large relative dielectric permittivity is desirable to maintain the same gate capacitance at larger dielectric layer thicknesses, which leads to a decrease in gate tunneling currents. Gate capacitance ($C_{gs}$) scales linearly with dielectric permittivity and the quantum mechanical Fowler-Nordheim tunneling current density ($J_{FN}$) of electrons decreases exponentially with increasing dielectric thickness. [1,2] Numerically, $C_{gs} = \varepsilon_r/t_d$, where $t_d$ is the dielectric thickness, and $J_{FN} = J_{FN,0} * e^{-F_0/F}$, where F is the electric field across the barrier and $F_0$ is the characteristic tunneling field, [3,4] which depends on the barrier height and barrier thickness. Accordingly, a thick gate dielectric with a higher dielectric permittivity (" high K", or high $\varepsilon_r$) offers the advantage of decreased electron tunneling and leakage currents without a decrease in gate capacitance. A commonly reported metric is equivalent oxide thickness, which is defined relative to SiO$_2$ dielectric thickness, with a low frequency $\varepsilon_r$ of ~3.9. Namely, a higher relative dielectric permittivity oxide material has a lower equivalent oxide thickness and is therefore attractive for vertical scaling of nitride high electron mobility transistors (HEMTs). Here, the term can be extended to nitride dielectrics like $Sc_xAl_{1-x}N$, as an "equivalent nitride thickness" with a comparison to a $Si_3N_4$ dielectric thickness with a low frequency $\varepsilon_r$ of ~7.5.

The vast majority of utilized dielectric materials in silicon electronics to replace $SiO_2$ [5,6] are oxides where the anion is oxygen. One of the reasons for this is due to the chemistry of oxide materials. Oxygen is strongly electronegative and tends to form ionic bonds with corresponding cations. Oxide materials tend to be air stable as nitrogen from the air tends to not displace oxygen from the crystal. In addition, the relatively small negative two (-2) charge of an oxygen anion in a crystal allows for a wide variety of possible crystal structures with charge neutrality to be possible. Combined, these are some of the reasons why a wide variety of compounds and bonding environments exist in the oxide family. Similarly, fewer chemical compounds and bonding environments currently exist in the nitride material family, due to a relatively large negative three (-3) anion charge and slightly lower electronegativity than oxygen. Accordingly, the search for nitrides as potential high K dielectric materials remains challenging. [7-10]

Most dielectric materials are deposited by *ex situ* deposition techniques and/or in a non-epitaxial manner. Current deposition of high-performance dielectrics on the III-nitride semiconductor electronic and photonic platform includes ex situ deposition of polycrystalline and/or amorphous $Al_2O_3$, $SiO_2$, and $Si_3N_4$. This requires air-exposure of potential active device layers, which can generate defective layers in-between the channel layer and dielectric via chemical modification, surface trap states, and other mechanisms. [11] Deposition of dielectric layers *in situ* can potentially reduce or eliminate these deleterious effects and relax the need for advanced surface preparation techniques related to those phenomena. To this end, the same epitaxial tools used to deposit AlN and GaN thin films, molecular beam epitaxy (MBE) and metal organic chemical vapor deposition (MOCVD), are ideal environments to deposit an *in situ*, epitaxial dielectric layers due to utilization of ultra-high vacuum conditions and high purity source materials in MBE and a chemically controlled environment in MOCVD.

To deposit an *in situ* dielectric layer on the III-nitride platform, one option is to deposit a nitride material with an enhanced relative dielectric permittivity. Sputter deposited $Sc_xAl_{1-x}N$ on metal electrodes has shown enhanced dielectric permittivity relative to AlN. [12-24] This is in addition to the several fundamental and technological advances that have been made in $Sc_xAl_{1-x}N$ via sputter deposition in recent years, [25-37] including viability as complementary metal oxide semiconductor (CMOS) compatible material due to the utilization of relatively low deposition temperatures and large wafer scale production. [38,39] Epitaxial growth of $Sc_xAl_{1-x}N$ via MBE and MOCVD on semiconducting layers such as GaN, AlN, and SiC has seen increased attention and development in recent years, [40-56] with enhanced piezoelectricity and even ferroelectric behavior seen in MBE films. [57-59] Notably, $Sc_xAl_{1-x}N$ is lattice-matched to the in-plane lattice parameter of GaN at ~x=0.18, making epitaxial stabilization of wurtzite $Sc_xAl_{1-x}N$ with tunable thicknesses on GaN possible. This is highly desirable as AlN grown on GaN forms cracks along the hexagonal unit cell axes at ~6-7 nm thickness to relieve the tensile misfit strain. [60] In contrast to sputter deposition, MBE typically produces high crystal quality, ultra-thin layers. Combined, these factors make MBE grown $Sc_xAl_{1-x}N$ an attractive dielectric material candidate, more specifically as an *in situ* epitaxial dielectric material with precise thickness control and near elimination of chemical contamination of heterojunction interfaces where mobile carriers reside in the transistor channels.

In this work, the structural and dielectric properties of epitaxial, single-crystalline $Sc_xAl_{1-x}N$/GaN heterostructures with Sc contents of 17-25% (x=0.17 to 0.25) are analyzed. The heterostructures are ~100 nm thick $Sc_xAl_{1-x}N$ grown on $n^+GaN$ ($N_d\sim2\times10^{19}/cm^3$) buffer layers on $n^+GaN$ bulk substrates. The resulting epitaxial $Sc_xAl_{1-x}N$ films have up to 2.4 times larger $\varepsilon_r$ values than AlN ($\varepsilon_r\sim8.5$) for 17-25% Sc (x=0.17-0.25). These properties indicate significant potential usage of $Sc_xAl_{1-x}N$ as an *in situ* epitaxial, single-crystalline dielectric that can be seamlessly integrated with the GaN electronics and photonics platform to boost device performance in the ultra-high speed radio frequency (RF) and mm wave transistors as well for high efficiency power electronics.

The $Sc_xAl_{1-x}N$/GaN heterostructures in this report were grown by MBE in a Veeco® GenXplor system with a base pressure of $10^{-10}$ Torr on Suzhou Nanonwin® 7x7 mm conductive n-type bulk GaN substrates. Sc metal source of nominally 99.9% purity (including C and O impurities) from Ames Laboratory was evaporated from a W crucible using a Telemark® electron beam evaporation system integrated with the MBE equipment. Flux feedback was achieved with an Inficon® electron impact emission spectroscopy (EIES) system by directly measuring the Sc atomic optical emission spectra. Aluminum (99.9999% purity), gallium (99.99999% purity), and silicon (99.9999% purity) were supplied using Knudsen effusion cells. Nitrogen (99.99995% purity) active species were supplied using a Veeco® RF UNI-Bulb plasma source, with growth pressure of approximately $10^{-5}$ Torr. The reported growth temperature is the substrate heater temperature measured by a thermocouple. In-situ monitoring of film growth was performed using a KSA Instruments reflection high energy electron diffraction (RHEED) apparatus with a Staib electron gun operating at 15 kV and 1.5 A. Post-growth X-Ray Diffraction (XRD) was performed on a PanAlytical Empyrean diffractometer at 45 kV, 40 mA with Cu Kα1 radiation (1.54057 Å). Post growth AFM measurements were performed using an Asylum Research Cypher ES system. Capacitance-voltage (CV) measurements were performed on a Cascade Microtech 11000 probe station in an $N_2$ ambient envrionment at 500 kHz AC frequency at room temperature on 40 μm and larger diameter circular Ti/Au electrodes patterned lithographically. Calibration for CV measurements consisted of an open circuit calibration with the probes floating, a short circuit calibration with probes on the same metal, and a 50 Ω impedance calibration with the probes across a 50 Ω resistor. CV measurements were performed using a parallel series conductance model to extract the capacitance of the system.

All $Sc_xAl_{1-x}N$/$n^+GaN$ heterostructures for this study were epitaxially grown in a reactive nitrogen environment in the MBE chamber at 200 W RF nitrogen plasma power and 1.95 standard cubic centimeters per minute (sccm) flow rate. The $n^+GaN$ layers were grown with a Si doping concentration of $\sim2\times10^{19}/cm^3$ under metal rich conditions with III/V ratio > 1 at 700 °C substrate temperature to promote smooth morphologies for the subsequent $Sc_xAl_{1-x}N$ layers. Sc and Al atomic percentages in the film were adjusted by the ratio of the respective fluxes from the effusion cell for Al and E-Beam for Sc. The $Sc_xAl_{1-x}N$ layers were grown under nitrogen rich conditions with III/V ratio ~ 0.85 at a substrate temperature of ~600 °C. Sc and Al were co deposited continuously during the growth. A more detailed description of the growth and justification for the Sc source usage is described elsewhere. [61]

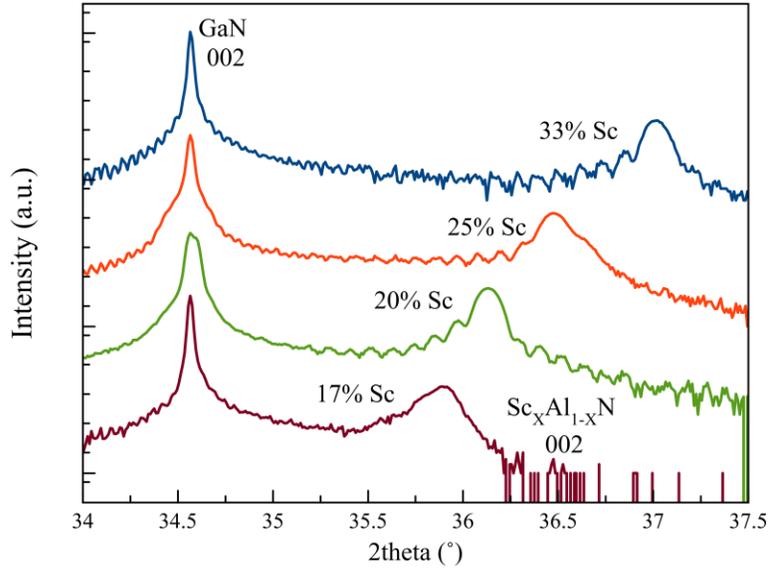

Fig. 1: Symmetric 2Θ-ω XRD 002 peak positions for ~100 nm thick $Sc_xAl_{1-x}N$ grown on $n^+$GaN bulk substrates. The 002 peak position increases as Sc content is increased. Sc compositions were obtained by comparing XRD peak positions to those of prior $Sc_xAl_{1-x}N$ samples calibrated by RBS.

To assess the crystallinity and epitaxial nature of the $Sc_xAl_{1-x}N$ films, post growth XRD measurements were performed, as shown in Figure 1. $Sc_xAl_{1-x}N$ peaks are observed, with an increasing 2θ position as Sc content is increased, indicating a decrease in c-axis lattice parameter. Interference fringes related to the thickness of the $Sc_xAl_{1-x}N$ layers and the coherent $Sc_xAl_{1-x}N$-GaN interface are visible, further indicative of epitaxial growth. The thickness of the $Sc_xAl_{1-x}N$ layer was verified by x-ray reflectivity (XRR) measurements (not shown). In situ RHEED images (not shown) confirmed the presence of epitaxial, single-crystalline growth of $Sc_xAl_{1-x}N$. Lattice parameter trends from XRD measurements are shown in Figure 2(a), with information from prior $Sc_xAl_{1-x}N$ samples grown at the same deposition conditions (e.g., III/V ratio, substrate temperature) calibrated by RBS measurements. Figure 2(b) shows the wurtzite crystal structure of AlN and the a and c-axis lattice parameters as a reference.

A non-monotonic trend of c-axis lattice parameter with Sc content is observed. A possible reason is the competition between an increasing average bond length as more Sc is incorporated into tetrahedral sites and structural distortion that tilts the tetrahedra away from the c-axis as more Sc is incorporated. [62,63] The increasing average bond length acts to increase the c-axis lattice parameter. The structural distortion that decreases the projection of the bonds onto the c-axis acts to decrease the c-axis lattice parameter. This non-monotonic behavior and deviation from Vegard's law is related to the fact that wurtzite ScN is not thermodynamically stable but rocksalt ScN is. [64-67] Accordingly, $Sc_xAl_{1-x}N$ can be viewed as a balance between competing tetrahedral coordination in wurtzite AlN and octahedral coordination in rocksalt ScN. The same behavior is also predicted to occur for $Y_xAl_{1-x}N$ [68] and when alloying other transition metals into AlN for similar reasons. This non-monotonic behavior is not observed in $Al_xGa_{1-x}N$ and $In_xAl_{1-x}N$ alloy systems as endmembers GaN and InN have thermodynamically stable wurtzite crystal structures.

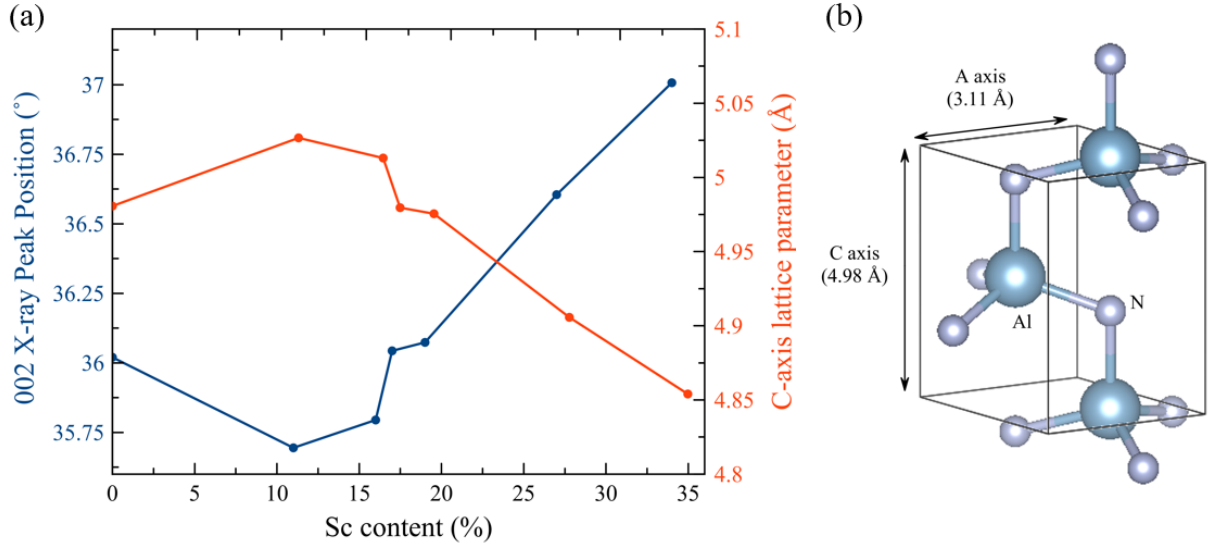

Fig. 2: (a) Symmetric 2Θ-ω XRD 002 peak positions for $Sc_xAl_{1-x}N$ and the corresponding calculated c-axis lattice parameters. The non-monotonic trend of c-axis lattice parameter with Sc content is observed. (b) Ball and stick crystal structure model of the wurtzite unit cell of AlN, showing the relative a and c axis orientations

Figure 3 shows CV results for the three Sc compositions studied in a capacitor geometry where the underlying $n^+GaN$ serves as an epitaxial bottom electrode. Top metal electrodes were patterned lithographically, and indium was adhered to the backside of the $n^+GaN$ substrate to serve as a bottom contact for vertical measurements. A relatively constant capacitance value with a small positive slope indicates uniform and high n type (donor) doping levels in the conductive $n^+GaN$ layer. In an ideal capacitor with electrodes and an insulating dielectric layer, the displacement current is 90 degrees out of phase with the applied voltage. This corresponds to a phase angle (θ) of 90 degrees. The phase angle is the angle between the circuit impedance and resistance (the real part of impedance). For an ideal capacitor, a parallel series-conductance model can be utilized to describe the equivalent circuit behavior of the system. For the three Sc compositions measured (17-25% Sc), the capacitance values are roughly constant across a large voltage range. The measured phase angle is close to 90 degrees for smaller voltage ranges and starts to deviate at larger biases due to the presence of conduction current from electrical leakage. As Sc content is increased, the voltage range in which the phase angle stays near 90 degrees decreases, indicating electrical leakage current increases as Sc content increases.

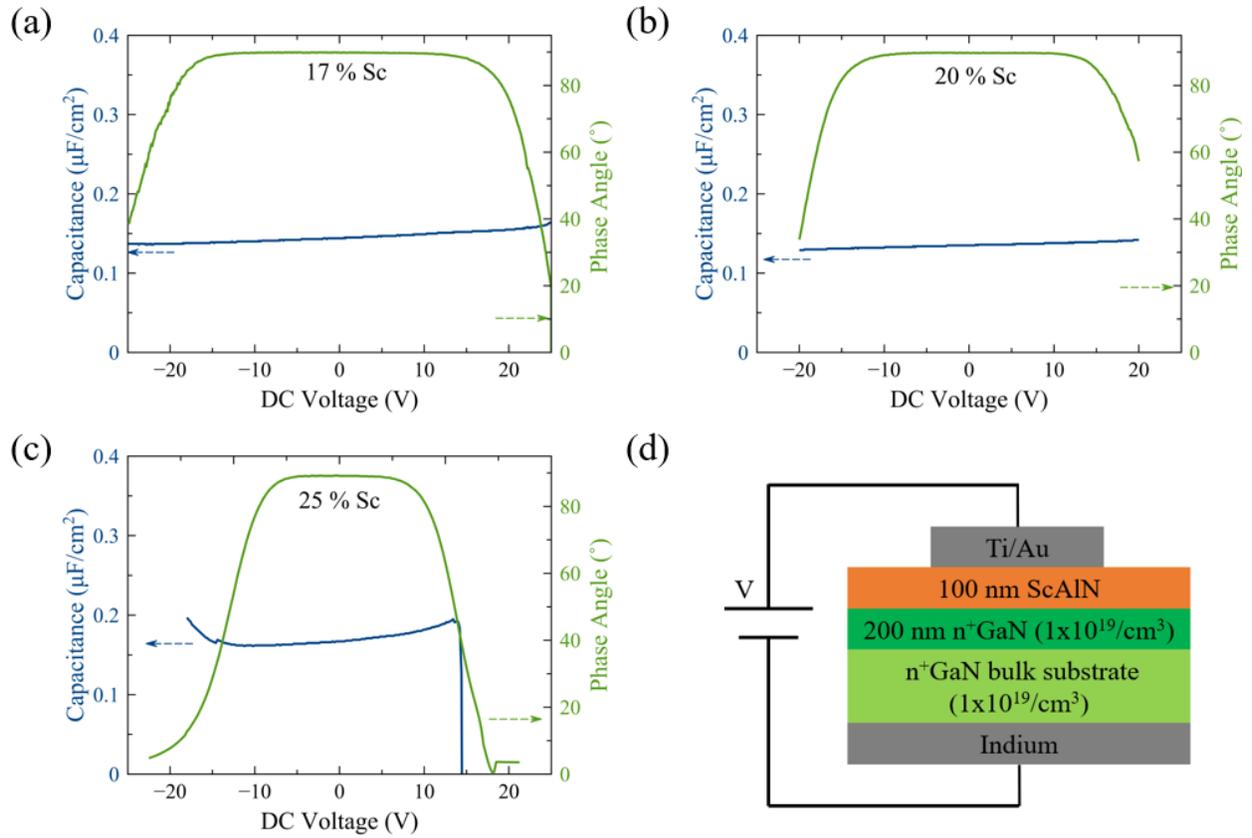

Fig. 3: (a-c) CV data at 500 kHz frequency showing capacitance per unit area and phase angle of Sc contents 17-25% (x=0.17-0.25) in $Sc_xAl_{1-x}N$ for 40 μm diameter circular electrodes. (d) Schematic of heterostructure configuration for CV measurements

A summary of CV measurement results for several Ti/Au electrode sizes is included in Table 1. The measured $\varepsilon_r$ values are relatively consistent across multiple electrode sizes, indicating the reliability of the measurements and the ability to probe large diameter electrodes without significant electrical leakage at low applied voltages. For 40 μm diameter electrodes, the nominal phase angles reported correspond to loss tangent (tanδ) values of 0.002, 0.003, and 0.016 for 17, 20, and 25% Sc, respectively. Delta (δ) is the angle between the circuit impedance and reactance (the imaginary part of impedance). These loss tangent values are competitive for potential dielectric materials but can be improved upon.

Table 1: Extracted dielectric permittivity ($\varepsilon_r$) for $Sc_xAl_{1-x}N$-$n^+GaN$ heterostructures and measured phase angle (°) at low voltages from CV measurements as a function of Sc content and metal electrode diameter (d). Measurements were conducted at 500 kHz frequency. Values for 33%Sc are not shown as the phase angle deviates significantly from 90° due to electrical leakage. Note the $\varepsilon_r$ for AlN is ~8.5.[69]

| Sc content | Parameters | d = 40 μm | d = 80 μm | d = 135 μm |
|---|---|---|---|---|
| 17 % | $\varepsilon_r$ | 16.8 | 15.6 | 15.1 |
|  | θ (°) | -89.85 | -89.95 | -89.60 |
| 20 % | $\varepsilon_r$ | 15.1 | 14.7 | 14.6 |
|  | θ (°) | -89.83 | -89.67 | -89.64 |
| 25 % | $\varepsilon_r$ | 21.6 | 21.0 | 20.9 |
|  | θ (°) | -89.06 | -89.09 | -88.65 |

The relative dielectric permittivity increases as Sc content is increased. From a simplified picture, this is expected as the total polarizability of the system increases when Sc is added. Sc, with a larger atomic radius than Al, has a larger electronic polarizability. Electronic polarizability increases as the atomic radius increases due to decreased nuclear shielding from the core electrons. Also, the ionic polarizability of $Sc^{+3}$ is larger than that of $Al^{+3}$, [70] partially due to the smaller electronegativity of $Sc^{+3}$. A smaller cation electronegativity implies the outermost shell has a weaker attractive force on neighboring anion electrons, which is in accordance with increased polarizability. Cation ionic polarizability is dependent on the specific crystal environment of the cation (e.g., unit cell volume, corresponding anion), so quantitative values are not easily determined. Nevertheless, the qualitative trends suggest that isoelectronic alloying of $Sc^{+3}$ on tetrahedral sites in the wurtzite unit cell will increase the polarizability of the system.

The mechanisms of dielectric polarization in order of increasing frequency cutoff are space charge, dipolar/orientational, ionic, and electronic. It is noted that space charge polarization refers to bulk or interfacial positive or negatively charged mobile carriers. For an inorganic single crystal like $Sc_xAl_{1-x}N$ without grain boundaries, mobile ions, and significant leakage from the metal electrodes at low voltages, space charge polarization is expected to be negligible. In addition, space charge polarization is generally too slow to follow the applied AC field at frequencies greater than 100 kHz. Ionic polarization and electronic polarization are present at 500 kHz and will persist up to much higher frequencies (GHz and beyond). Dipolar/orientational polarization is not expected for inorganic, non-ferroelectric materials where the bonds are not free to rotate significantly. However, this phenomenon can exist in ferroelectric materials while dipoles attempt to align with the applied AC electric field. The measured $\varepsilon_r$ values in this report are similar to those obtained in sputter deposited $Sc_xAl_{1-x}N$ films and for those predicted by theory. [12-24,71] Thus, the values obtained in this report are expected to be representative of the low frequency $\varepsilon_r$ values for the $Sc_xAl_{1-x}N$ materials system.

Figure 4 shows a comparison between the low frequency relative dielectric permittivity of common III-nitride semiconductors and dielectrics used for III-nitride electronic and photonic devices. $Sc_xAl_{1-x}N$ (x=0.25) has an $\varepsilon_r$ value that is the highest of any existing III-nitride material ($\varepsilon_r \sim 21$) and comparable with other dielectrics such as $HfO_2$ and $Ta_2O_5$. Though the $\varepsilon_r$ value for InN is ~15, the relatively low bandgap and conductive nature of InN has precluded its use as a dielectric material, which gives merit to $Sc_xAl_{1-x}N$ as a potential nitride dielectric alternative to $Si_3N_4$. Separate films of $Sc_xAl_{1-x}N$ (x=~0.20) of ~200 nm thickness grown on $AlN$-$Al_2O_3$ substrates with the same conditions (e.g., III/V ratio, temperature) as those grown on $n^+GaN$ show a near band-edge absorption of ~5.40 eV (not shown) from absorption measurements. Optical absorption measurements could not be performed directly on the $Sc_xAl_{1-x}N$-GaN heterostructures due to the smaller bandgap and concomitant absorption of GaN. The bandgaps of $Sc_xAl_{1-x}N$ on AlN seen here are similar to prior reports for sputter deposited and MBE grown $Sc_xAl_{1-x}N$ at a similar Sc content. [39,50,72] Assuming the bandgap of $Sc_xAl_{1-x}N$ is the same when grown on AlN and GaN, this gives $Sc_xAl_{1-x}N$ a relatively large bandgap and relative dielectric permittivity. Also, $Sc_xAl_{1-x}N$-GaN two-dimensional electron gases (2DEGs) show strong quantum confinement, indicating favorable band offsets that are desirable. These combined factors give merit towards the usage of $Sc_xAl_{1-x}N$ as a potential high K gate dielectric material. As $Sc_xAl_{1-x}N$ development continues and the material system becomes more insulating and can sustain large electric fields, its promise as a dielectric material will expand significantly as well.

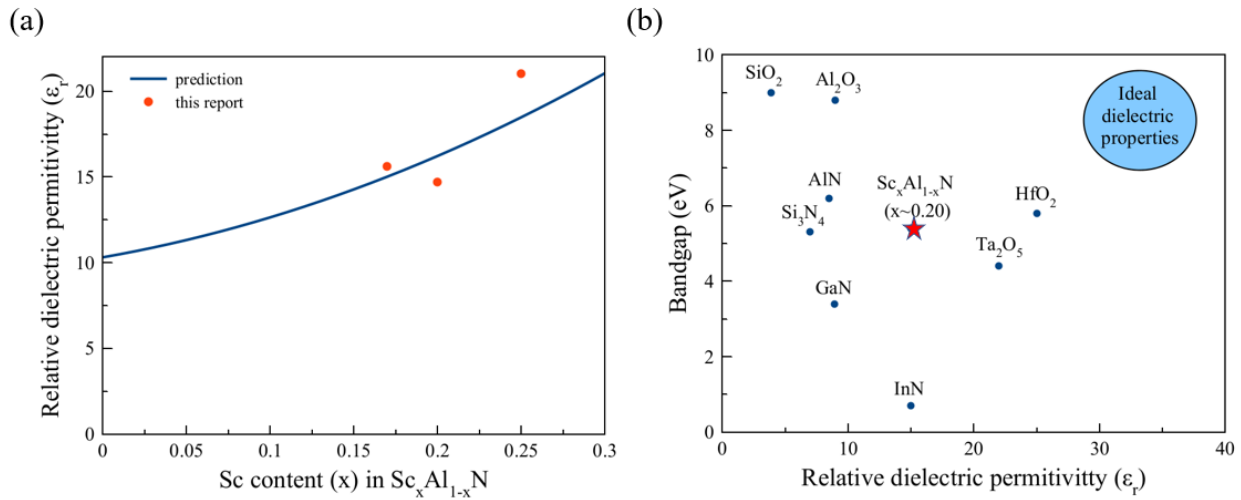

Fig. 4: (a) Prediction of low frequency $\varepsilon_r$ values for various Sc contents in $Sc_xAl_{1-x}N$ in epitaxial pseudomorphic heterostructures from [71] with overlaid data for the Sc compositions included in this study. (b) Comparison of low frequency relative dielectric permittivity and optical bandgap of common III-nitride semiconductors and dielectric materials utilized for III-nitride materials. Data taken from [73]. The results for ~20% Sc $Sc_xAl_{1-x}N$ in this report are highlighted with a red star. The $Sc_xAl_{1-x}N$ bandgap value used is from separate $Sc_xAl_{1-x}N$ samples grown on AlN-$Al_2O_3$ substrates.

In conclusion, insights into the lattice parameter evolution of $Sc_xAl_{1-x}N$ grown epitaxially on GaN is obtained, with a trend of decreasing c-axis lattice parameter values once Sc content is increased past the nominally lattice-matched composition of ~18% Sc (x=0.18). The relative dielectric permittivity increases as Sc content is increased, reaching a value of ~21 at ~25% Sc (x=0.25), a value that is competitive with some existing "high k" dielectric materials. The realization of an enhanced $\varepsilon_r$ in $Sc_xAl_{1-x}N$ grown by MBE on GaN paves the way for the potential usage of $Sc_xAl_{1-x}N$ as an *in situ* dielectric material that can utilize the novel properties and functionality of the $Sc_xAl_{1-x}N$ materials system. This can lead to increased performance of nitride HEMTs by increasing the breakdown voltage and decreasing tunneling current density through reducing the electric field due to the "high K" properties of $Sc_xAl_{1-x}N$.

This work was supported by the DARPA Tunable Ferroelectric Nitrides (TUFEN) program monitored by Dr. Ronald G. Polcawich. The work was also supported in part by NSF DMREF grant 1534303, Cornell's nanoscale facility (grant NCCI-1542081), AFOSR grant FA9550-20-1-0148, NSF DMR-1710298, and the Cornell Center for Materials Research Shared Facilities which are supported through the NSF MRSEC program (DMR-1719875). The authors would like to acknowledge the Materials Preparation Center, Ames Laboratory, US DOE Basic Energy Sciences, Ames, IA, USA for supplying the Sc source material. The authors would also like to acknowledge Professor Darrell Schlom for helpful discussions.

The data that support the findings of this study are available from the corresponding author upon reasonable request.